\documentclass[conference]{IEEEtran}

\usepackage{cite}
\usepackage{amsmath,amssymb,amsfonts}
\usepackage{graphicx}
\usepackage{textcomp}
\usepackage{xcolor}
\usepackage{soul}
\usepackage{tabularx}
\usepackage{multirow}
\usepackage{cite}
\usepackage{hyperref}

\usepackage{algpseudocode}
\usepackage{algorithm}
\usepackage{xspace}
\usepackage{amsmath}
\usepackage{upgreek}
\newtheorem{definition}{Definition}
\newcommand{\prob}{\mathbb{P}}

\newcommand{\A}{\ensuremath{\mathcal{A}}}

\usepackage{threeparttable}  
\usepackage{booktabs}
\usepackage{multirow}
\usepackage{bm}

\algdef{SE}[DOWHILE]{Do}{doWhile}{\algorithmicdo}[1]{\algorithmicwhile\ #1}%

\newcommand{\dpsgd}{DP-SGD}
\newcommand{\ewtune}{\ensuremath{\texttt{EW\mbox{-}Tune}}}

\def\BibTeX{{\rm B\kern-.05em{\sc i\kern-.025em b}\kern-.08em
    T\kern-.1667em\lower.7ex\hbox{E}\kern-.125emX}}

\title{Privately Fine-Tuning Large Language Models with Differential Privacy}

 \makeatletter
\newcommand{\linebreakand}{%
  \end{@IEEEauthorhalign}
  \hfill\mbox{}\par
  \mbox{}\hfill\begin{@IEEEauthorhalign}
}
\makeatother

\author{\IEEEauthorblockN{Rouzbeh Behnia$^* $\thanks{$^* $Equally contributing authors (alphabetically ordered by the last name.)}}
\IEEEauthorblockA{School of Information Systems\\ and Management \\ 
University of South Florida\\
     Sarasota, USA \\
    behnia@usf.edu}
\and
\IEEEauthorblockN{Mohammadreza (Reza) Ebrahimi$^* $}
\IEEEauthorblockA{School of Information Systems\\ and Management \\ 
University of South Florida\\
   Tampa, USA \\
 ebrahimim@usf.edu}
\and
\IEEEauthorblockN{Jason Pacheco}
\IEEEauthorblockA{Department of Computer Science\\
    {University of Arizona}\\
    Tucson, USA\\
    pachecoj@cs.arizona.edu}
\linebreakand 
\IEEEauthorblockN{Balaji Padmanabhan}
\IEEEauthorblockA{School of Information Systems\\ and Management \\
    {University of South Florida}\\
    Tampa, USA\\
    bp@usf.edu}
}

\IEEEoverridecommandlockouts

\begin{document}

\maketitle
  \thispagestyle{plain}
  \pagestyle{plain}

\begin{abstract}
Pre-trained Large Language Models (LLMs) are an integral part of modern AI that have led to breakthrough performances in complex AI tasks. Major AI companies with expensive infrastructures are able to develop and train these large models with billions and millions of parameters from scratch. Third parties, researchers, and practitioners are increasingly adopting these pre-trained models and fine-tuning them on their private data to accomplish their downstream AI tasks. However, it has been shown that an adversary can extract/reconstruct the exact training samples from these LLMs, which can lead to revealing personally identifiable information. The issue has raised deep concerns about the privacy of LLMs. Differential privacy (DP) provides a rigorous framework that allows adding noise in the process of training or fine-tuning LLMs such that extracting the training data becomes infeasible (i.e., with a cryptographically small success probability). While the theoretical privacy guarantees offered in most extant studies assume learning models from scratch through many training iterations in an asymptotic setting, this assumption does not hold in fine-tuning scenarios in which the number of training iterations is significantly smaller. To address the gap, we present \ewtune, a DP framework for fine-tuning LLMs based on Edgeworth accountant with finite-sample privacy guarantees. Our results across four well-established natural language understanding (NLU) tasks show that while \ewtune~adds privacy guarantees to LLM fine-tuning process, it directly contributes to decreasing the induced noise to up to  5.6\% and improves the state-of-the-art LLMs performance by up to 1.1\% across all NLU tasks. We have open-sourced our implementations for wide adoption and public testing purposes.
\end{abstract}


\section{Introduction}

Large language models (LLMs) have become an integral component of modern AI.
Deep learning architectures with billions of parameters are often designed based on transformers, a building block first introduced by Google’s BERT \cite{devlin2018bert}.
LLMs provide breakthrough performance in complex AI tasks
such as dialogue systems \cite{ham2020end} and text/automated story generation \cite{fang2021transformer}. Being equipped with the hardware infrastructure, major AI companies such as Open AI and Facebook provide new LLMs trained on the public data from the Internet \cite{brown2020language,radford2019language}. Common examples include, RoBERTa \cite{brown2020language} and GPT \cite{radford2019language}. RoBERTa's training dataset includes English Wikipedia and millions of online news crawled from the internet. Similarly, GPT was trained on outbound links from Reddit.

AI researchers and practitioners often fine-tune these pre-trained models on their downstream AI tasks using their own  private data to accomplish downstream tasks such as malware detection \cite{hu2021single}, text-to-image generation \cite{ramesh2021zero}. However, recently, it has been shown that these pre-trained models are vulnerable to privacy attacks \cite{CarliniTWJHLRBS21}. This problem is mainly due to the model's tendency to memorize training samples without overfitting, also known as the ''memorization issue" \cite{carlini2019secret}. This issue could lead to three major types of privacy attacks: membership inference, model inversion, and training data extraction. 
\begin{itemize}
    \item Membership inference \cite{hisamoto2020membership}: determines whether a certain user’s data was included in the training.
    \item Model inversion \cite{fredrikson2015model}: approximate the reconstruction of the training data.
    \item Training data extraction \cite{CarliniTWJHLRBS21}: aims to exactly reveal the training samples which makes this type of attack the most powerful one with the most adverse consequences for the users.
\end{itemize}

While all three types of attacks can jeopardize the privacy of the users whose information is in the training data, training data extraction  directly targets users’ personally identifiable information and can endanger users’ identity via revealing important information such as their address, social security number, phone number, etc. The fine-tuned LLMs used by third parties on their private data will face the same privacy concerns. These privacy concerns around the issue necessitate privacy-preserving approaches for fine-tuning LLMs. Such an approach will allow third parties to privately fine-tune the LLMs on their private data without any information leak about their private training samples.

Differential Privacy (DP) is a promising approach to ensure the training data privacy with theoretical guarantees \cite{AbadiCGMMT016}. DP provides a mathematically rigorous framework with privacy guarantees that enables Stochastic Gradient Descent (SGD), the cornerstone of learning in LLMs, in a private setting. In such a setting, SGD can be applied as a randomized \textit{mechanism} multiple times in each iteration of the training.
Most DP methods provide asymptotic guarantees. For theoretical guarantees, the number of SGD applications (known as compositions) is often assumed to be unlimited in most privacy studies. This assumption leads to asymptotic guarantees in these studies (i.e., infinite compositions of SGD in the limit). However, in LLM fine-tuning the number of SGD iterations is not only limited but also quite small (i.e., in the order of several thousand) \cite{yu2021differentially}.


In this study, through a DP lens, and thanks to the finite sample guarantee achieved by Edgeworth expansion \cite{WangEW22}, we propose a novel LLM fine-tuning framework, called  \ewtune, with finite-sample guarantees. \ewtune~operates based on an effective DP accounting approach known as Edgeworth accountant, proposed in \cite{WangEW22}. Edgeworth accountant computes the amount of noise that is required to be added to the gradients in SGD to guarantee a certain privacy budget (see Section~\ref{dplit}). \ewtune~also leverages the latest efficient reparametrization technique proposed in \cite{yu2021lowrank}. 

\subsection{Our contribution}
While \ewtune~is a general framework, we showcase its performance by  focusing on its application to enhance the privacy of LLM during fine-tuning.
Our contribution to the LLM's private fine-tuning is two-fold:
\begin{itemize}
\item Our study serves as the first step towards fine-tuning LLMs in a differentially private setting when the number of compositions (i.e., the applications of differentially private SGD) is finite and limited to only several thousand  (less than 4,000 times in our experiments).   Compared to the existing methods that provide an asymptotic bound on the privacy budget, through utilizing Edgeworth accountant, \ewtune~is able to provide a non-asymptotic privacy bound by using Berry-Esseen bound derived from the Edgeworth approximation. In the case of fine-tuning LLMs, given the finite number of compositions, for the same privacy budget, \ewtune~induces less noise to SGD compared to the state-of-the-art. This directly improves the learning and the accuracy of the model.

\item It is known that while fine-tuning via DP enhances the model's privacy, it can negatively affect the model's utility (i.e., performance) \cite{AbadiCGMMT016}. Our experiments show that \ewtune~significantly contributes to the state of the art by enhancing the privacy of LLMs while preserving their utility/accuracy compared to multiple recent alternative methods across several important downstream benchmark tasks including text classification, entailment detection, and question answering.  Overall, \ewtune~decreases the noise-induced to SGD up to 5.6\%. \ewtune~also enhances the state-of-the-art model’s accuracy by up to 1.1\%.
\end{itemize}


\section{Background and Related Work}

We review three areas of the literature: (1) LLMs to identify the state-of-the-art in language modeling and their fine-tuning. (2) Differentially private deep learning as the overarching framework to rigorously guarantee the privacy of fine-tuning LLMs. (3) Edgeworth accountant as an emerging accountant method that provides fine-sample guarantees, which could be a useful tool for fine-tuning LLMs.

\subsection{Large Language Models (LLMs)} Large language models are deep neural network architectures with billions of parameters \cite{liu2019roberta,baevski2020wav2vec,raffel2020exploring}. They often benefit from an encoder-decoder architecture that generates high-quality representations from sequence data (text, image, malware, genes, etc.). Most LLMs use specific types of layers with self-attention mechanisms known as transformers to dynamically assign weights to input elements based on their surrounding context \cite{liu2019roberta}. Transformers enable LLMs to provide high-quality representations of the input sequence. At a high level, LLMs can be categorized into two types: masked and autoregressive. 

Masked language models are trained to predict a masked token based on its surroundings. Highly effective examples of masked language models include BERT \cite{devlin2018bert} and RoBERTa \cite{liu2019roberta}. On the contrary, autoregressive language models learn to predict the next token based on the previously generated ones, which makes them suitable for text generation tasks \cite{brown2020language,yang2019xlnet}.

Due to their ability to produce high-quality representations from input, masked language models are widely used in major downstream AI tasks including text classification, question answering, semantic entailment detection, and speech recognition.

 Pre-trained LLMs are often fine-tuned on specific tasks and datasets, through which  the weights of the original model are updated to better tune for the domain-specific data and task in hand.

\subsection{Differentially Private Deep Learning}
\label{dplit}
 Differential privacy  \cite{DworkMNS06}, formally defined in Definition \ref{Def:DP}, computes a privacy guarantee when the results of an algorithm, run on private data, are made public. When applied to machine learning, a differentially private (DP)  mechanism allows for the public release of the model parameters while ensuring the privacy of the original training data. 

\definition{A randomized mechanism $M: \mathcal{X} \rightarrow \mathcal{Y}$  is $(\epsilon,\delta)$-DP, if for all adjacent datasets $X, X' \in \mathcal{X}$, differing in a single element only, and all $Y \subset \mathcal{Y}$, $  \prob(M(X)\in Y)\leq e^\epsilon\prob(M(X')\in Y) + \delta$ holds. 
}\label{Def:DP}

In Definition \ref{Def:DP}, $(\epsilon,\delta)$ is often referred to as the privacy budget. $\epsilon$ ~defines the distance between the two sides of the equation and $\delta$ defines a failure probability. 
Differential privacy enjoys from properties such as robustness to auxiliary information and composability. The former guarantees privacy even with the emergence of new side-information to the adversary and the latter allows for modular design of mechanisms. Essentially, composability implies that if two mechanisms $M_1(\cdot)$ and $M_2(\cdot)$  are DP, and $M(X) = (M_1(X), M_2(X))$, then $M$ is also differentially private. 

\noindent\textbf{Differentially private Stochastic Gradient Descent (\dpsgd).} The gold standard to achieve differential privacy in deep learning is to update the neural network parameters with \emph{noisy} gradients. This is achieved by a randomized algorithm called \dpsgd~\cite{AbadiCGMMT016} via the two following steps: 
\begin{itemize}
    \item[--] \textbf{Gradient clipping:} given a clipping norm $C$, the gradient of each sample $x$, $\mathbf{g}(x)$, is clipped $\mathbf{g}'(x)\gets \mathbf{g}(x)/\max(1,\frac{\|\mathbf{g}(x)\|_2}{C}) $. 
    \item[--] \textbf{Gaussian mechanism:}  the clipped gradients are aggregated and then an isotropic Gaussian noise from $\mathcal{N}(0,C^2\sigma^2)$,  with $\sigma$ as a  noise multiplier, will be added to the gradients.

\end{itemize}
The noise multiplier in \dpsgd~is determined by the privacy budget $(\epsilon,\delta)$, the number of training rounds $m$, and the sampling probability $q=B/N$ for batch size $B$ and $N$ as the total number of samples.

In their seminal work, Abadi et al. \cite{AbadiCGMMT016} introduced a method called Moments accountant (MA) for computing an upper bound for the privacy curve of compositions of DP algorithms. The method was then used to track the privacy loss of \dpsgd~by computing the \emph{privacy curve} for each training iteration with itself $m$ times, where $m$ is the total number of iterations.
In \cite{RDP-Mironov17}, the MA framework is instantiated with Renyi Differential Privacy (RDP). However, these algorithms, while being efficient (runtime is independent of $m$),  provide an upper bound which is rather impractical. 

The Gaussian Differential Privacy (GDP) framework \cite{GDP-Bu,GDP-Jinshuo} also called $f$-DP is devised based on the central limit theorem (CLT). The GDP framework offers a nice characterization of differential privacy using  hypothesis testing interpretation \cite{kairouz2015composition}. GDP can only provide an \emph{approximation} for the privacy curve and it was shown to underreport the true epsilon value in \cite{GopiLW21}.

Using the notion of privacy loss random variables (PRV) \cite{DworkR16}, Meiser and Mohammadi \cite{MM18} introduced an algorithm called \emph{privacy bucket} for approximately composing privacy curves. By employing the notion of PRV, one can utilize the nice property of PRV to compute the composition of $m$ mechanisms $M = M_1 \circ M_2 \circ \cdots \circ M_m $ by simply summing their corresponding PRVs $\mathcal{D} = \sum_{i=1}^{m}\mathcal{D}_i$. The distribution of $\mathcal{D}$ can then be approximated by computing the convolution of its underlying distributions $\mathcal{D}_1, \dots, \mathcal{D}_m$. Koskela et al. \cite{KoskelaJH20} used fast Fourier transform (FFT) to compute the convolution efficiently. Following \cite{KoskelaJH20}, Gopi et al. \cite{GopiLW21} leveraged FFT to  numerically compose trade-off functions. Their accountant, called the PRV accountant addresses the underestimation of $f$-DP and provides  an upper-bound and lower-bound on the leakage of $\epsilon$.

\subsection{Edgeworth Accountant}\label{sec:ewacc}

As noted, at the heart of $\ewtune$~is Edgeworth accountant \cite{WangEW22}. Edgeworth accountant relies on $f$-DP \cite{GDP-Jinshuo}, which as discussed above, offers a full characterization of differential privacy by utilizing hypothesis testing interpretation. Informally, differential privacy  measures the hardness in distinguishing any pair of (neighboring) datasets based on the information obtained from a mechanism $M$. In \cite{GDP-Jinshuo}, the authors formulated the notion of indistinguishably as a hypothesis testing problem for two neighboring datasets $S$ and $S'$. Therefore, the hypotheses are formed as $H_0$: the underlying dataset is $S$ and $H_1$: the underlying dataset is $S'$, where the output of $M$ is the basis for conducting a hypothesis testing problem. Let $P$ and $Q$ denote the probability distributions of $M(S)$ and $M(S')$, respectively. Now, for a rejection rule $0\leq \o\leq 1 $, and hypotheses $H_0:P$ and $H_1:Q$, the trade-off function  $f= T(P,Q)(\alpha)=\inf\{\beta_{\o}:\alpha_{\o} \leq \alpha\}$ defines the mapping from the Type-I error to Type-II error, where $\alpha_{\o}=\mathbb{E}_P[\o]$ and  $\beta_{\o}=1-\mathbb{E}_Q[\o]$. To compute the composition of the trade-off functions of the form $f=\bigotimes^m_{i=1}f_i$, let us realize the $i$-th composition by two hypothesis $H_{0,i}=w_i\sim P_i $  and  $H_{1,i}=w_i\sim Q_i$. Now, to evaluate the trade-off function $f=\bigotimes^m_{i=1}f_i$, we distinguish between two composite hypothesis $H_{0,i}=\mathbf{w}\sim P_1\times P_2\times\cdots\times P_m $  and  $H_{1,i}=\mathbf{w}\sim Q_1\times Q_2\times\cdots\times Q_m $ for $\mathbf{w}=(w_1,\dots,w_m)$.

Edgeworth accountant \cite{WangEW22} defines random variables called privacy-loss log-likelihood ratios (PLLRs) to enable the lossless conversion of the $f$-DP guarantee into a collection of $(\epsilon,\delta)$-DP  guarantees.  PLLRs are defined as a Radon-Nikodym derivatives of the hypotheses above as $X_i\equiv\log\frac{dQ_i(\zeta_i)}{dP_i(\zeta_i)}$ and $Y_i\equiv\log\frac{dQ_i(\chi_i)}{dP_i(\chi_i)}$ for $\zeta \sim P_i$ and $\chi \sim Q_i$. The authors in \cite{WangEW22} showed the primal-dual relationship between $f$-DP and a collection of $(\epsilon,\delta(\epsilon))$-DP via $\delta = 1-F_{Y,m}(\epsilon)  - \mathbf{e}^\epsilon(1-F_{X,m}(\epsilon))$ where $F_{X,m}$ and $F_{Y,m}$ are the CDFs of $\sum_{i=1}^m X_i$ and $\sum_{i=1}^m Y_i$, respectively. To compute PLLRs through composite hypotheses, the Edgeworth accountant uses a family of PLLR sequences to compose the tightest possible trade-off function that satisfies all $f^{(\alpha)}$-DP.


Assuming that for each $\alpha$, one can find a series of PLLRs corresponding to $f^{(\alpha)}$, we can compute a collection of $(\epsilon,\delta^{(\alpha)}(\epsilon))$-DP guarantee\footnote{The authors in \cite{WangEW22} discuss how one can invert the equation to compute a $\epsilon$ for a given $\delta$}. Then one has  to compute an approximate CDF as a random variable $X=\sum_{i=1}^m X_i$ using Edgeworth expansion to  output the Edgeworth accountant approximates as $F_{X,m}$ and $F_{Y,m}$.

\section{Proposed Method}
\subsection{Threat Model}

\noindent\textbf{Adversary capabilities and objectives}. We consider an adversary $\A$ to have black-box access to the language model.  In this work, following \cite{CarliniTWJHLRBS21}, we assume that \A~does not have access to the model's specific weights and hidden states but is able to obtain next-word predictions and compute the probability of arbitrary sequences for instance via access to auto-complete models. The ultimate goal of the adversary is to extract (the memorized) training data from the model. The severity of an attack is increased if more examples could be extracted from the model. 
 
 \begin{figure}[t]
\centering
\includegraphics[width=\columnwidth]{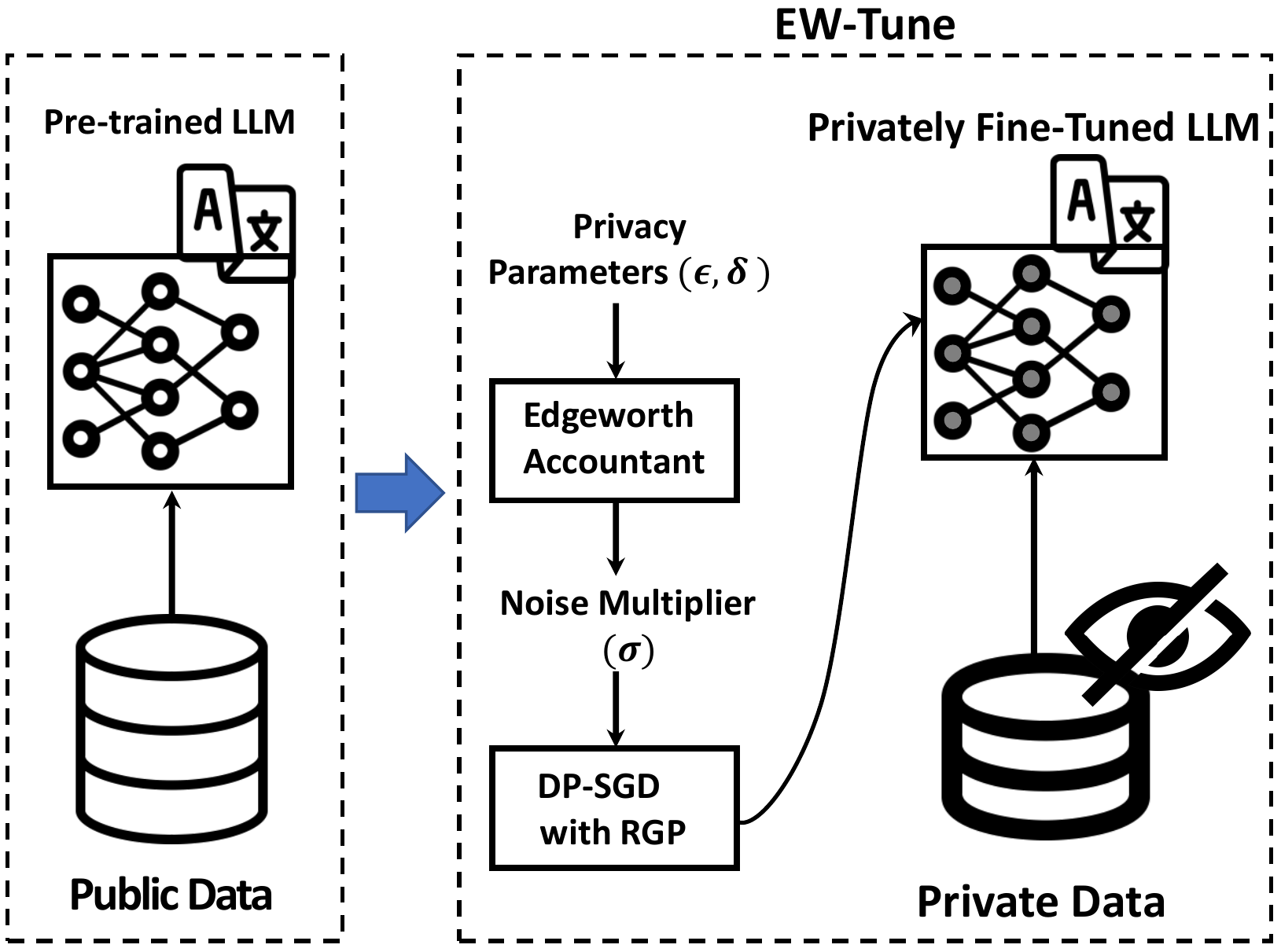}\vspace{-17pt}
\caption{Abstract View of the Proposed \ewtune~Framework}
\label{ew}
\end{figure}
\noindent\textbf{Adversary's target and tasks}. 
While \ewtune~is a general framework that can be applied to enhance the privacy of any LLM during fine-tuning. For specificity, we focus on one of the most highly-adopted masked language models in AI tasks, a successor of Google's BERT, named roBERTa \cite{liu2019roberta}. roBERTa mainly owes its popularity  to its ability to learn the bidirectional representation of the sentence. These high-quality representations that are not available in autoregressive language models such as GPT, specifically contribute to breakthrough results in common downstream natural language understanding (NLU) tasks such as sentiment analysis and text categorization. Also, to show the generalizability of \ewtune, we will test its utility and privacy guarantees across four important and complex NLU tasks (all included in the well-known General Language Understanding Evaluation (GLUE) benchmark dataset \cite{glue_2019}). Each task is associated with a well-established dataset:
\begin{itemize}
    \item MNLI \cite{williams_mnli_2018}: The Multi-Genre Natural Language Inference (MNLI) is a collection of 433,000 sentence pairs that are annotated with semantic entailment information \cite{williams_mnli_2018}. The LLM's task in this corpus is to identify the semantic relationships between a given pair of sentences (entailment, contradiction, or neutral relationship).
    \item QNLI \cite{glue_2019}: The Question-answering Naural Language Inference (QNLI) is a natural language inference dataset collected from Wikipedia that consists of 110,400 question-paragraph pairs, where only one of the sentences in the paragraph is the answer to the corresponding question. The LLM's task is to determine whether a sentence includes the answer to a given question. 
    \item QQP \cite{glue_2019}: The Quora Question Pairs (QQP) dataset includes over 400,000 question pairs. Each question pair is annotated to indicate whether these questions are semantically equivalent (i.e., paraphrase of each other). The LLM's task is to determine whether either of the questions is the paraphrase of the other one.
    \item SST-2 \cite{sst_2013recursive}: The Stanford Sentiment Treebank (SST-2) includes 68,800 sentences from movie reviews and annotations of their sentiment. The LLM's task is to predict the sentiment (positive or negative) of a given sentence.
\end{itemize}

To operationalize defense against the above threat model, we propose \ewtune, a general framework for fine-tuning LLMs for different downstream tasks. Figure~\ref{ew} depicts the components of our proposed \ewtune \hspace{1pt} framework.


\begin{algorithm}\caption{\ewtune~Framework}\label{alg:ewTune}

\begin{algorithmic}[1]

\State \textbf{Input:} Examples $\{x_i\}_{i=1}^N$, mechanisms $\{M_i\}_{i=1}^m$, a weight matrices $\{\mathbf{W}^{(l)}\}^H_{l=1}$, warm-up steps $T_w$, group size $B$, gradient clipping bound $C$, the failure probability $\delta$ and an  initial privacy budget $\epsilon$ and a $k$-th order Edgeworth expansion.
\State Given the sampling probability $q=B/N$, for a given $\delta$, for each mechanism and all $\alpha$ encode the PLLRs $[(X_i^\alpha,Y_i^\alpha)]$  and the cumulants up to order $k+2$  
\State For each $\alpha$ compute the Edgeworth approximation and calculate $\epsilon^{(\alpha)}(\delta)$ and  supremum $\sup_\alpha \epsilon^{(\alpha)} (\delta)$
 
\State Given an $\epsilon_{init}$ and an arbitrary initial $\sigma_{arb}$ (e.g., $\sigma =10$)  \While {($\epsilon < \epsilon_{init} $ AND $  r < \sigma$)}  
\State Recompute $\epsilon$ (Steps 2-4) and  refine and reduce $\sigma$ using an initial reducing factor $r$ (e.g., $r=0.5$)
\EndWhile  
 \State Choose a gradient-carrier matrix $\{\mathbf{W}^{(l)}\}^H_{l=1}$ according to  \cite{yu2021lowrank}.
 \State Sample a batch of examples with probability $q$.
 \State Compute historical update to find gradient carrier and decompose 
  and compute the low-rank gradient carrier $\mathbf{L}$ and $\mathbf{R}$ via \cite[Algorithm 2]{yu2021lowrank}  
  \State Run  reparametrized forward/backward process and compute individual gradients  $\{\partial_i\mathbf{L}_t^{(l)},\partial_i\mathbf{R}_t^{(l)}\}_{l\in[H],i\in S_t}$

  \State Given $C$ and $\sigma$, clip and add noise to the individual gradients as in Section \ref{dplit} to get $\Tilde{\partial}\mathbf{L}_t^{(l)} $ and $ \Tilde{\partial}\mathbf{R}_t^{(l)}$
  
  \State Construct $\Tilde{\partial}\mathbf{W}_t^{(l)}= (\Tilde{\partial}\mathbf{L})\mathbf{R}+\mathbf{L}(\Tilde{\partial}\mathbf{R})- \mathbf{L}\mathbf{L}^T(\Tilde{\partial}\mathbf{L})\mathbf{R}$

  \State \textbf{Output:} $\langle\Tilde{\partial}\mathbf{W}_t^{(l)}, (\epsilon,\delta)\rangle$
\end{algorithmic}

\end{algorithm}
 \begin{table*}[!h]
\centering
\caption{Performance Comparison of \ewtune against state of the art DP methods across four different NLU tasks conducted by roBERTa LLM}
\renewcommand{\arraystretch}{}
\begin{tabular}{|c||c|c|c|c||c|c|c|c|}
\hline
\multirow{2}{*}{Method} & MNLI  & QNLI & QQP   & SST-2 & MNLI  & QNLI & QQP   & SST-2\\    
\cline{2-9}
&\multicolumn{4}{c||}{Accuracy} & \multicolumn{4}{c|}{Noise Multiplier}\\
\hline
RDP& 81.25\%& 86.63\%&84.60\% & 89.24\% &0.65 &0.829 &0.6575 & 0.921\\ \hline
PRV &  81.22\% & 86.79\% &84.78\% &  91.82\% &0.607 &0.768 &0.6135 &0.8485 \\ \hline
\ewtune & \textbf{81.81}\% & \textbf{87.71}\% & \textbf{84.91}\%& \textbf{92.19}\% & \textbf{0.573}& \textbf{0.739}&\textbf{0.579} &\textbf{0.8215} \\ \hline
\end{tabular}
\begin{tablenotes}
\centering
\footnotesize
{ \item \footnotesize Performance reported for $\delta = 1e-6$ for MNLI, QNLI, and QQP; $\delta = 1e-5$ for SST-2; batch size $=2000$ }
\end{tablenotes}
\label{comparison} 
\end{table*}
As seen in Figure~\ref{ew}, the pre-trained LLM is learned on a public dataset (e.g., internet) from scratch by a major AI company (e.g., Google and OpenAI) (shown in the left side of Figure~\ref{ew}). Subsequently, the pre-trained model is used as an input to \ewtune~to fine-tune with privacy guarantees expressed by privacy parameters (i.e., privacy budget $\epsilon \in [0,\infty)$ and failure probability $\delta \in [0,1]$) (as shown in the right side of Figure~\ref{ew}). The privacy parameters $(\epsilon, \delta)$ are provided by the user/practitioner. A smaller $\epsilon$ indicates better privacy preservation (and lower utility/performance). $\delta$ denotes the probability that the training examples are accidentally being leaked. In the context of LLM, a suitable value  for $\epsilon$ is between 5 to 8, and $\delta$ is recommended to take a value in the order of  the inverse of the size of the training samples \cite{yu2021differentially}. 

After the user provides the privacy parameters, the Edgeworth accountant algorithm \cite{WangEW22} is utilized to (1) compute the number of compositions (i.e., applications of DP-SGD) for a given dataset. (2) compute the amount of noise that guarantees the given privacy budget. Subsequently, any variation of DP-SGD \cite{AbadiCGMMT016,yu2021lowrank}, as described in Section~\ref{dplit}, can be used to fine-tune the LLM on the private dataset based on the appropriate noise multiplier $\sigma$ obtained in the previous step. Due to its breakthrough performance, we utilized a recent version of the DP-SGD algorithm that is based on a new method called  reparameterized gradient perturbation (RGP) \cite{yu2021lowrank}.  In the original DP-SGD \cite{AbadiCGMMT016}, the noise introduced highly depends on the model parameters, and the per-example gradient clipping results in very high memory and computation overhead. RGP addresses the problems of \dpsgd~by reparameterizing each layer's weight matrix $\mathbf{W}$ into two low ranks gradient-carrier matrices $\mathbf{L}$ and $\mathbf{R}$ and a residual weight matrix $\Tilde{\mathbf{W}}$. Finally, all the transformer layers of the LLM will be fine-tuned on the private data through DP-SGD. The output of the \ewtune~framework is the fine-tuned LLM as shown in the right side of Figure~\ref{ew}.

Algorithm \ref{alg:ewTune} presents a more detailed version of our framework (\ewtune). The algorithm starts by computing an $\epsilon$ for a given $\delta$ based on the Edgeworth accountant explained in Section \ref{sec:ewacc} (Steps 2-4) by first computing the PLLRs and then approximating their CDF using Edgeworth expansion. Then we compute $\epsilon^{(\alpha)}(\delta)$ and  its supremum. Next, in Steps 5-8, we compute our final  $\epsilon$ and an appropriate noise multiplier $\sigma$  to be used in our reparametrized gradient perturbation (i.e., noise addition as in Section \ref{dplit}). In Step 5, with each iteration of the loop we compute $\sigma  = \sigma - r$  and the initial tuning factor $r$ is reduced by a constant factor (e.g., we use $r/10$ in our code). Finally, given the $\sigma$,  the RGP algorithm efficiently perturbs the updated parameters (Steps 9-14). For each update with the weight matrix $\mathbf{W}$, the algorithm works in four main steps. In the first step the gradient carrier matrices $\mathbf{L} $  and $\mathbf{R}$ are generated  via the decomposition method proposed in \cite[Algorithm 2]{yu2021lowrank}. The output of this step is the orthonormalize version (via Gram-Schmidt orthonormalization process) of the gradient carrier matrices. Next, the  weight matrices are reparametrized to compute  and store individual gradients via the forward/backward process presented in \cite[Section 2]{yu2021lowrank}. In the third step, the gradients are clipped and made noisy similar to the \dpsgd~method presented in Section \ref{dplit}. Lastly, in Step 14, the noisy aggregated gradients of the carrier matrices $\langle\Tilde{\partial}\mathbf{W}_t^{(l)}$ are used to compute the gradients of the original weights.

\section{Experiments}
We have open-sourced our optimized framework for wide adoption and testing purposes at the following link.

 \vspace{2mm}
\begin{center}
	{\url{https://github.com/star-ailab/LLM_Tune}}
\end{center} 

\subsection{Experiment Setup}

\ewtune~was developed and run on a single NVIDIA GeForce RTX 3090 with 10,496 CUDA cores and 24 GB of internal memory. To ensure that the LLM loading and fine-tuning can take place in the internal memory we selected the roBERTa.base pre-trained model with 125 million parameters. This LLM can be accessed from \url{https://github.com/facebookresearch/fairseq/tree/main/examples/roberta}.

\subsubsection {Parameter settings} Consistent with \cite{glue_2019}, we set the train and test partition for each dataset: (MNLI: 393,000 for training and 20,000 for testing; QNLI: 105,000 for training and 5400 for testing; QQP: 364,000 for training and 391,000 for testing, SST-2: 67,000 for training and 1800 for testing). 
 
To facilitate comparison, we set the privacy parameters $\epsilon$ and $\delta$ consistent with \cite{yu2021differentially}. Accordingly, we set $\epsilon=8$, $\delta=1e-6$ for larger datasets (i.e., MNLI, QNLI, and QQP; each with several hundred thousand of samples) and $\delta=1e-5$ for the smaller dataset (i.e., SST-2; with tens of thousands of samples). 

\subsubsection{Benchmark Experiments} Following \cite{WangEW22,yu2021differentially} We evaluated the performance of the proposed \ewtune~framework against two widely-used state-of-the-art DP alternatives: Renyi Differential Privacy (RDP) \cite{RDP-Mironov17,AbadiCGMMT016} and Privacy Loss Random Variables (PRV) \cite{GopiLW21}. To rigorously evaluate \ewtune, we conduct two sets of experiments (Section \ref{results}).

In Experiment 1, we evaluate the accuracy of the LLM in solving the four mentioned NLU tasks (MNLI, QNLI, QQP, SST-2) after fine-tuning  with \ewtune,  RDP and PRV. In Experiment 2, we evaluate the amount of noise induced by alternative privacy accountant algorithms to that of \ewtune~for different values of $\epsilon$.

\subsection{Results}\label{results}
\subsubsection{Experiment 1}
Table~\ref{comparison} shows the results of comparing the accuracy and noise multiplier of the proposed \ewtune \hspace{1pt} against RDP and PRV across four NLU datasets (MNLI, QNLI, QQP, and SST-2) carried out by the roBERTa LLM. To report the performance, we  repeated each experiment 3 times and reported the average.  
The highest accuracy in performing each task appears in bold font. As seen in Table~\ref{comparison}, \ewtune~yields the highest performance (81.81\% on MNLI, 87.71\% on QNLI, 84.91\% on QQP, and 92.19\% on SST-2) among its counterparts. At the heart of \ewtune, the Edgeworth accountant utilizes an accurate privacy computation method called $f$-differential privacy ($f$-DP)  along with  Edgeworth approximation, in place of CLT, that enjoys from a much better convergence rate. This allows \ewtune~to achieve the privacy guarantees by applying less noise to the transformer layers of LLMs during training. The noise multiplier (i.e., the standard deviation of Gaussian noise distribution) is shown on the right side of Tabel~\ref{comparison}. As shown in Tabel~\ref{comparison}, \ewtune~yields the lowest noise multiplier. More specifically, as compared to the state-of-the-art, the noise multiplier is up to 4\%, 3.2\%, 5.6\%, and 3.2\%   lower (for $5 \leq  \epsilon\leq  8$) for MNLI, QNLI, QQP, and SST-2, respectively. In Tabel~\ref{comparison}, the highest performance numbers and lowest noise multipliers are indicated in boldface. We note that if instead of utilizing \dpsgd~with RGP \cite{yu2021lowrank}, one were to instantiate these LLM's with the original \dpsgd~\cite{AbadiCGMMT016}, the smaller noise multiplier numbers in \ewtune~would result in a much higher performance gap with our counterparts. 


\subsubsection{Experiment 2}
\begin{figure}[t]\label{noise}
\centering
\includegraphics[width=\columnwidth]{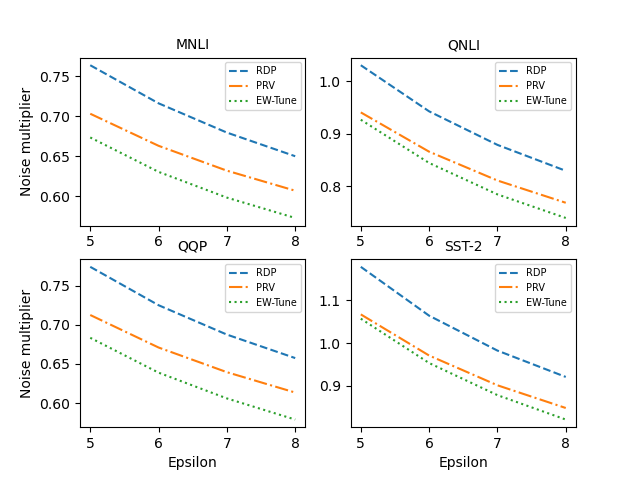}\vspace{-8mm}
\caption{The Changes of Noise Multiplier based on different values of $\epsilon$ across three Privacy Accounting Algorithms (RDP, PRV, and \ewtune.)}
\vspace{-3mm}
\end{figure}

Experiment 2 evaluates the amount of noise induced by \ewtune~and other benchmark privacy accountant algorithms at $\delta=1e-6$ for MNLI, QNLI, QQP, and $\delta=1e-5$ for SST-2. As noted, it is desirable to achieve the same privacy budget ($\epsilon$) by applying less noise to the transformer layers of the LLM during the fine-tuning. As shown in Figure~\ref{noise}, \ewtune~yields the lowest amount of noise across the benchmark privacy accounting methods (i.e., RDP and PRV). As shown in Figure~\ref{noise}, when $\epsilon$ changes from 5 to 8, \ewtune~induces the lowest amount of noise into the SGD process for all four NLU tasks (MNLI, QNLI, QQP, and SST-2).  

Overall, The results of Experiments 1 and 2 on four complex NLU task shows that \ewtune~is able to enhance the performance thorough applying less noise to the SGD process, while achieving the same privacy budgets as its counterpart algorithms. \ewtune~outperforms the other privacy accountant methods for different values of privacy budget.
\section{Conclusion and Future Work}
In this work we  presented a new framework called \ewtune, specifically designed for fine-tuning LLMs. By utilizing the state-of-the-art privacy accountant and gradient perturbation methods, \ewtune~is able to provide finite-sample privacy guarantee by introducing less noise as compared to the existing methods. \ewtune~introduces up to 6\% less noise when privately training large language models which contributes to up to 1.1\% performance improvement. This can contribute to addressing the gap in privacy and accuracy trade-off in the realm of data privacy and AI. 

An interesting future work would be to further study the relationship between the introduced noise and training accuracy by focusing on the model's total number of parameters, dataset size, task objectives, and the number of compositions.


\section*{Acknowledgment}
We would like to thank Hua Wang from the Statistics Department at University of Pennsylvania for illuminating discussions on Edgeworth accountant and helpful comments on its implementation.

\bibliography{PPML}
\bibliographystyle{IEEEtran}

\vspace{12pt}

\end{document}